\documentclass[conference]{IEEEtran}
\IEEEoverridecommandlockouts
\usepackage{cite}
\usepackage{amsmath,amssymb,amsfonts}
\usepackage{algorithmic}
\usepackage{graphicx}
\usepackage{textcomp}
\usepackage{xcolor}
\usepackage{subcaption}
\usepackage{hyperref}
\usepackage{multirow}

\def\BibTeX{{\rm B\kern-.05em{\sc i\kern-.025em b}\kern-.08em
    T\kern-.1667em\lower.7ex\hbox{E}\kern-.125emX}}

\usepackage{comment}
\newcommand*{\X}{\mathbf{X}_{ft}}
\newcommand*{\Xe}{\mathbf{\hat{X}}_{ft}}

\newcommand*{\gkts}{g_{kts}}
\newcommand*{\zso}{z_{so}}
\newcommand*{\Wfo}{\mathbf{W}_{fo}}
\newcommand*{\Hfs}{\mathbf{H}_{fs}}

\begin{document}
% \title{Multichannel Singing Voice Separation using DOA Constrained CNMF model informed by MaD TwinNet
\title{Multichannel Singing Voice Separation by Deep Neural Network Informed DOA Constrained CNMF\vspace{-25pt}
\thanks{Part of the research leading to these results has received funding from the European Research Council under the European Union’s H2020 Framework Programme through ERC Grant Agreement 637422 EVERYSOUND.}
}
\makeatother   
\author{\IEEEauthorblockN{Antonio J. Mu\~noz-Montoro, Julio J. Carabias-Orti}
\IEEEauthorblockA{\textit{Departamento de Telecomunicaci\'{o}n} \\
\textit{Universidad de Ja\'{e}n}\\
Ja\'{e}n, Spain \\
\{jmontoro, carabias\}@ujaen.es}
\and
\IEEEauthorblockN{Archontis Politis, Konstantinos Drossos}
\IEEEauthorblockA{\textit{Audio Research Group} \\
\textit{Tampere University}\\
Tampere, Finland \\
\{firstname.lastname\}@tuni.fi}
% \and
% \IEEEauthorblockN{Julio Carabias Orti}
% \IEEEauthorblockA{\textit{Departamento de Telecomunicaci\'{o}n} \\
% \textit{Universidad de Ja\'{e}n}\\
% Ja\'{e}n, Spain \\
% carabias@ujaen.es}
}
\IEEEaftertitletext{\vspace{-2\baselineskip}}
\maketitle
\begin{abstract}
This work addresses the problem of multichannel source separation combining two powerful approaches, multichannel spectral factorization with recent monophonic deep-learning (DL) based spectrum inference. Individual source spectra at different channels are estimated with a Masker-Denoiser Twin Network (MaD TwinNet), able to model long-term temporal patterns of a musical piece. The monophonic source spectrograms are used within a spatial covariance mixing model based on Complex Non-Negative Matrix Factorization (CNMF) that predicts the spatial characteristics of each source. The proposed framework is evaluated on the task of singing voice separation with a large multichannel dataset. Experimental results show that our joint DL+CNMF method outperforms both the individual monophonic DL-based separation and the multichannel CNMF baseline methods.
\end{abstract}

% \vspace{-8pt}

%%%%%%%%%%%%%%%%%%%%%%%%%%%%%%%%%%%%%%%%%%%%%%%%%%%%%%%%%%%%%
%%%%%%%%%%%%%%%%%%%%%%%%%%%%%%%%%%%%%%%%%%%%%%%%%%%%%%%%%%%%%
\section{Introduction}

% Singing voice separation refers to the decomposition of a music recording into two tracks, the singing voice on one side, and the instrumental accompaniment on the other side. 
Singing voice separation refers to the decomposition of a musical mixture into two parts; the singing voice and the instrumental accompaniment. 
This audio signal processing task is useful in many applications such as karaoke, speaker identification, speaker-specific information retrieval, word recognition, and others. 
Classical algorithms in the literature extract the singing pitch from song mixtures as the cue for subsequent separation~\cite{hsu2012tandem,durrieu2011musically}.
% Classical algorithms in the literature extract the singing pitch from song mixtures as the cue for subsequent separation~\cite{li2007separation,hsu2012tandem, virtanen2008combining,durrieu2011musically}.
% When training data is available, spectral patterns can be learned in advance providing reliable results when the testing signals do not differ significantly from the training conditions~\cite{raj2007separating,ozerov2007adaptation}. 
% When training data are available, spectral patterns can be learned in advance providing reliable results when the testing signals do not differ significantly from the training conditions~\cite{raj2007separating}.
With sufficient amount of data, spectral patterns can be learned in advance, providing reliable results when the testing signals do not differ significantly from the training conditions~\cite{raj2007separating}. 

% Recent approaches are based on Deep Learning (DL)~\cite{purwins2019deep}, commonly estimating a time-frequency mask for each source in the mixture together with a soft-masking based strategy~\cite{huang2015,grais2016single,Chandna2017}.
Recent approaches are based on deep learning (DL), commonly estimating a time-frequency mask for each source in the mixture together with a soft-masking based strategy~\cite{Chandna2017}. To overcome the limitations of training from pre-computed source masks~\cite{mimilakis2017recurrent}, feature extraction together with separation can be implicitly incorporated into the network architecture, for example by using an end-to-end convolutional neural network~\cite{stoller2018wave}. 
% To overcome the limitations of training from pre-computed source masks~\cite{mimilakis2017recurrent}, feature extraction together with separation can be implicitly incorporated into the network architecture, for example by using an end-to-end convolutional neural network (CNN)~\cite{pascual2017segan,stoller2018wave}. 
% The separation is performed by passing the encoder features to DNNs whereas the reconstruction of the waveforms is achieved by inverting the encoder operation. 
Recently, a proposal based on denoising autoencoders~\cite{mimilakis2018monaural} using a Masker-Denoiser (MaD) architecture % that employs RNNs of stochastic depth in order to recover the time-frequency mask. 
%The usage of RNNs allows for efficient modeling of longer time dependencies of the input data. 
and further extended in~\cite{Drossos2018MaDSeparation} with a regularization technique called TwinNet, provided state-of-the-art results on monaural singing voice separation. Alternatively to autoencoders, deep-clustering mask estimation has been also applied to the same task~\cite{luo2017deep}.
%
% When multichannel signals are available, separation can be improved by taking into account the spatial locations of sources or the mixing process. Multichannel-NMF approaches model the latent source magnitude- or power-spectrograms with NMF while the spatial mixing system is modeled without the non-negativity constraint~\cite{Ozerov2012,Sawada2013,Nikunen2014,Carabias-Orti2018}. 

When multichannel signals are available, separation can be improved by taking into account the spatial locations of sources or the mixing process. Multichannel non-negative matrix factorization (NMF) based approaches model the latent source magnitude- or power-spectrograms with NMF while the spatial mixing system is modeled without the non-negativity constraint~\cite{Sawada2013,Nikunen2014,Carabias-Orti2018}. 
%%%%%%%%%%%%%%%%CHANGED by Antonio%%%%%%%%%%%%%%%%
The spatial properties of the sources can be modeled using a spatial covariance matrix (SCM) which encodes magnitude and phase differences between the recorded channels. Authors in~\cite{Sawada2013} proposed to estimate unconstrained SCM mixing filters together with NMF magnitude model to identify and separate repetitive frequency patterns corresponding to a single spatial location. To mitigate the effect of the spatial aliasing, Nikunen and Virtanen~\cite{Nikunen2014} proposed a SCM model based on DOA kernels to estimate the inter-microphone time delay given a looking direction. Carabias et al.~\cite{Carabias-Orti2018} proposed a SCM kernel based model where the mixing filter is decomposed into two direction dependent SCMs to represent  and estimate disjointly both time and level differences between array channels. The main drawback of these strategies is the large number of parameters which have to be tuned and thus, without any prior information, these methods are prone to converge to local minima, especially in reverberant environments.
%%%%%%%%%%%%%%%%%%%%%%%%%%%%%%%%%%%%%%%%%%%%%%%%
% Recent works have tried to exploit multichannel audio with DNN-based approaches. Deep-clustering approaches are augmented with spatial information in~\cite{wang2018multi, drude2019integration} with large improvements over monophonic versions, 
Recent works have tried to exploit multichannel audio with deep neural network (DNN) based approaches. 
% Deep-clustering methods are augmented with spatial information in~\cite{drude2019integration} with large improvements over monophonic versions, while the proposals in~\cite{Nugraha2016, seki2019generalized} combine DNN-based source spectrogram estimation with MNMF-inspired spatial models.
Deep-clustering methods are augmented with spatial information in~\cite{drude2019integration} with large improvements over monophonic versions, while the proposals in~\cite{Nugraha2016} combine DNN-based source spectrogram estimation with MNMF-inspired spatial models. 
Finally, a fully spatio-spectral factorization DNN is proposed as deep tensor factorization in~\cite{casebeer2019deep}.
%Recently, the NMF spectrogram model has been replaced by deep learning strategies, and use of deep neural networks (DNNs) for modeling the source spectrogram in combination with Gaussian SCM model was proposed in~\cite{Nugraha2016} where it was reported to outperform NMF-based models.

In this work, we propose a joint DNN-CNMF method for multichannel singing voice separation. We exploit the strong monophonic separation performance of DNN masking-based approaches using the MaDTwinNet method~\cite{Drossos2018MaDSeparation}. However, the estimated source spectra are integrated spatially with the complex-NMF (CNMF) model with spatial covariances modeling inter-channel dependencies. Directional information is modeled similarly to~\cite{Nikunen2014}, as a weighted combination of direction of arrival (DoA) kernels. However, contrary to the blind source separation CNMF of~\cite{Nikunen2014,Carabias-Orti2018}, knowledge of the source spectra as delivered by MaDTwinNet alleviates the underperformance of such approaches under realistic reverberant conditions~\cite{Carabias-Orti2018}. Additionally, it is shown that imposing the CNMF spatial model across channels on the initial separated source spectra, improves further separation and signal quality over the DNN-based monophonic separation.
%The use of RNNs for singing voice is motivated by the strong and long-term temporal structures which are inherent in music signals.
%Then, the mixing model is estimated using a Spatial Covariance CNMF model where the spatial location is modeled as a weighted combination of DoA kernels~\cite{Nikunen2014}. Opposite to the beamforming-like CNMF model in~\cite{Nikunen2014,Carabias-Orti2018} where the sources are obtained by grouping time-frequency components arriving from certain spatial location, in this work, the prior estimation of the sound source allows to estimate the source spatial location and the mixing properties alleviating the underperformance of the CNMF based methods under realistic reverberant conditions~\cite{Carabias-Orti2018a}.

%%%%%%%%%%%%%%%%%%%%%%%%%%%%%%%%%%%%%%%%%%%%%%%%%%%%%%%%%%%%%
%%%%%%%%%%%%%%%%%%%%%%%%%%%%%%%%%%%%%%%%%%%%%%%%%%%%%%%%%%%%%
\section{Monophonic Separation Based on DNNs}
A challenging case of singing voice separation is when the input mixture is monophonic, referred to as monaural singing voice separation (MSVS). Approaches for MSVS can vary, depending on the targeted output of the DNN and the calculation of the loss for optimizing the DNN~\cite{Drossos2018MaDSeparation,mimilakis:2018:icassp}. Some approaches predict directly the targeted source magnitude spectrogram, others estimate the time-frequency masks, while some more recent ones combine the previous by estimating masks, but optimizing with a source spectra distance loss ~\cite{weninger:2014:globalsip,mimilakis:2020:taslp}. Additionally, approaches that predict high-dimensional embeddings that need to be further clustered to result in usable separation masks, such as deep clustering~\cite{luo2017deep}, have gained popularity. 
%The same concept of optimizing the parameters follow the approaches in the second group, where the input is the mixture but the targeted output is a pre-computed mask that will perform the separation. That is, the optimization of the parameters of the DNN is performed according to the distance of the predicted and targeted masks. Recently~\cite{mimilakis:2017:mlsp,mimilakis:2018:icassp,Drossos2018MaDSeparation} is proposed another approach for MSVS, where the optimization of the parameters of the DNN is performed using the result of the Hadamard product of the output of the DNN and its input. That is, the output of the DNN is multiplied element-wise with its input and the result is considered as the predicted output, which is used in order to optimize the parameters of the DNN. This approach yields state-of-the-art results, has been employed before MSVS in different fields, and termed as skip-filtering connections in MSVS and signal approximation method in its original proposal~\cite{weninger:2014:globalsip,mimilakis:2020:taslp}. 

Our employed DNN-based method for MSVS is named MaDTwinNet~\cite{Drossos2018MaDSeparation}, is based on RNNs, accepts as an input the magnitude spectrum of a monophonic music mixture, and outputs the predicted magnitude spectrogram of the targeted source. MaDTwinNet utilizes a two-step process; firstly the separation of the target source (termed as masking), and secondly the enhancement of the predicted source (termed as denoising). Additionally, MaDTwinNet employs a method for enhancing the learning of long temporal structures with RNNs, which is termed TwinNets~\cite{serdyuk:2017:iclr}. More specifically, the masking process in MaDTwinNet is implemented by the Masker, which is an RNN-based auto-encoder structure, using residual connections between layers of the encoder, and a trainable affine transform at the end. Masker accepts as an input the magnitude spectrum of a monophonic mixture % $\mathbf{V}\in\mathbb{R}^{T\times F}_{\geq0}$, 
$x_{ft}\in\mathbb{R}_{\geq0}$, having $T$ vectors of $F$ frequency bins, and outputs $\phi_{fts}\in\mathbb{R}_{\geq0}$, $S$ being the number of sources. The estimated magnitude spectrum $\hat{y}_{fts}'\in\mathbb{R}_{\geq0}$ of the $s$-th targeted source is obtained by
%\begin{equation}\label{eq:masker-madtwinnet}
%    \hat{\mathbf{V}}'^{j} = \tilde{\mathbf{M}}\odot\mathbf{V}\text{ ,}
%\end{equation}
\begin{equation}\label{eq:masker-madtwinnet}
    \hat{y}'_{fts} = \phi_{fts} x_{ft}\text{ ,}
\end{equation}
\noindent
The process described in Eq.~\eqref{eq:masker-madtwinnet} is expected to introduce artifacts in the predicted magnitude spectrum $\hat{y}'_{fts}$. For that reason, MaDTwinNet employs a second module, termed as Denoiser, which takes as an input $\hat{y}'_{fts}$, outputs $\psi_{fts}\in\mathbb{R}_{\geq0}$, and produces an enhanced (i.e. cleaned/denoised) version of it as
\begin{equation}
    \hat{y}_{fts} = \psi_{fts} \hat{y}'_{fts}\text{ .}
\end{equation}
\noindent
Additionally, MaDTwinNet employs TwinNets as a regularization of the Masker, in order to enhance the learning capabilities of the RNNs in the Masker. This results in a training process where the employed RNNs in the Masker can model better the long-term temporal patterns in the input magnitude spectrum~\cite{serdyuk:2017:iclr,Drossos2018MaDSeparation}. Finally, MaDTwinNet is optimized by minimizing the loss
\begin{equation}
    \mathcal{L} = \mathcal{L}_{M} + \mathcal{L}_{D} + \mathcal{L}_{TwinNet} + \lambda\Xi\text{ ,}
\end{equation}
\noindent
where $\mathcal{L}_{M}=D_{\text{KL}}(y_{fts}||\hat{y}'_{fts})$ and $\mathcal{L}_{D}=D_{\text{KL}}(y_{fts}||\hat{y}_{fts})$ are the losses for the Masker and Denoiser, respectively, $D_{\text{KL}}$ is the generalized Kullback-Leibler divergence, $\mathcal{L}_{TwinNet}$ is the total loss associated with the TwinNets, and $\Xi$ is the total regularizing terms employed, scaled by $\lambda$.

\section{Proposed CNMF with MaDTwinNet}
\begin{figure}[!t]
\centering
\includegraphics[width=0.485\textwidth]{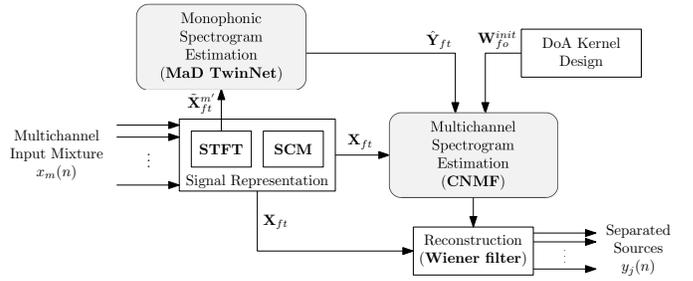}
\caption{Block diagram of the proposed system.}
\label{fig:diagram}
\end{figure}
In this work we propose a joint multichannel singing voice separation strategy using DNNs and a SCM-based CNMF signal model. In particular, the source spectrogram is estimated using the DNN-based MaDTwinNet method~\cite{Drossos2018MaDSeparation}. The output of MaDTwinNet is used in the proposed multichannel CNMF method to learn the mixing filter. Finally, the source separation is refined attending to the spatial properties learned for each source. The block diagram of the proposed framework is depicted in Fig.~\ref{fig:diagram}.

First, the SCM signal representation is computed from the STFT of the multichannel mixture. Then, the magnitude spectrogram of the singing voice is estimated independently for each channel using MaDTwinNet~\cite{Drossos2018MaDSeparation}. The use of this method is motivated since it provides state-of-the-art results for the task of the single channel singing voice separation, as demonstrated in the Signal Separation Evaluation Campaign (SiSEC 2018)~\cite{Stoter2018}. Then, the estimated source spectrogram in the predominant channel is used as prior information in the proposed SCM-based CNMF framework. The proposed method consist in two steps: 1) Estimation of the mixing filter of the multichannel input signal using the output spectrograms of the network, 2) Refinement of the source magnitude spectrograms accounting to the spatial properties learned in step 1. Finally, a generalized soft-masking strategy is used to obtain the source reconstruction.

% \subsection{Multichannel Mixing Model}\label{sec:mixingmodel}
\subsection{Multichannel mixing model and filter estimation using CNMF with DoA kernels}\label{sec:mixingmodel}
% Estimating the mixing parameters using CNMF methods with absolute phase information~\cite{Kameoka2009ComplexSignals, Rodriguez-Serrano2016a} can be a cumbersome task due to phase being generally stochastic and dependent on features such as sound onsets.
Estimating the mixing parameters using CNMF methods with absolute phase information~\cite{Rodriguez-Serrano2016a} can be a cumbersome task due to phase being generally stochastic and dependent on features such as sound onsets.
Alternatively, we account for the inter-channel phase differences between microphones using a spatial covariance matrix (SCM) signal representation as in~\cite{Carabias-Orti2018a,Nikunen2014}. In particular, the SCM mixing model can be expressed as
\begin{equation}\label{eq:mixingmodel}
    \X \approx \Xe = \sum_{s=1}^S \mathbf{H}_{fs} \hat{y}_{fts}\text{,}
\end{equation}
\noindent
where $\X \in \mathbb{C}^{M \times M}$ is the SCM representation of the $M$ channels input signal for each time-frequency point $(f,t)$, $\hat{y}_{fts}$ denotes the magnitude spectrogram for each source $s \in [1, ..., S]$ and $\mathbf{H}_{fs} \in \mathbb{C}^{M \times M}$ is the SCM representation of the spatial frequency response. As explained in \cite{Nikunen2014}, under the assumption that the sources are uncorrelated and sparse (i.e. only a single source is active at each time frequency $(f,t)$ point), the SCM model in Eq.~\eqref{eq:mixingmodel} can be approximated to be purely additive. However, this assumption does not always hold in reverberant environments.
%
% \subsection{Mixing Filter estimation using CNMF with DoA Kernels}\label{sec:SSmodel}

The source mixing filter $\mathbf{H}_{fs}$ in Eq.~\eqref{eq:mixingmodel} models both amplitude and phase differences between channels, however it has no explicit relation to spatial locations and, without further constraints, the dimensionality of the optimization problem many cause inaccurate solutions \cite{Sawada2013}. 
Beamforming-inspired SCM methods in \cite{Nikunen2014} model the SCM mixing filter as a weighted combination of DoA Kernels pointing towards a certain number of spatial directions. The source separation is then performed by grouping NMF components according to their spatial direction. This method performs well under anechoic or moderate reverberant conditions. However, without further information about the sources or the room response, the separation quality degrades in typical reverberant music recording conditions.
To mitigate this problem for the task of singing voice separation, we propose to use prior information from the sources obtained by the MaDTwinNet method. In particular, the proposed SCM-based signal model using the mixing model in~\cite{Nikunen2014} can be expressed as 
\begin{equation}\label{eq:signalmixingmodel}
    \X \approx \Xe = \sum_{s=1}^S \underbrace{\sum_{o=1}^O \Wfo \zso}_{\Hfs} \hat{y}_{fts}\text{,}
\end{equation}
\noindent
where $\hat{y}_{fts}$ is the time-frequency spectrogram obtained using the MaDTwinNet method~\cite{Drossos2018MaDSeparation}. 
Note that we applied MaDTwinNet independently for each channel and we use as prior information the source estimation with higher energy across channels. 
As proposed in \cite{Nikunen2014} the SCM mixing filter $\mathbf{H}_{fs}$ can be decomposed as a linear combination of the DoA kernels $\mathbf{W}_{fo} \in \mathbb{C}^{M \times M}$ multiplied by the spatial weights $z_{so}\in\mathbb{R}_{\geq0}$, which relates sources $s$ with spatial directions $o$. As in \cite{Nikunen2014}, the DOA kernels $\mathbf{W}_{fo}$ are initialized a priori for every spatial direction $o$. 
Regarding the magnitude spectrogram $\hat{y}_{fts}$, two types of sound sources are considered for the task of singing voice separation, i.e. singing voice and music accompaniment. 

For the estimation of the mixing filter parameters of Eq.~\eqref{eq:signalmixingmodel} we use the  majorization-minimization algorithm proposed in~\cite{Sawada2013,Carabias-Orti2018}. Using this approach, the cost function can be described using both Euclidean or Itakura Saito (IS) divergence. In this work, we use the IS divergence because it provides better separation results as explained in \cite{Carabias-Orti2018}. For the sake of brevity, we will directly present here the update rules. In the case of the spatial weights $\zso$, the update rules are given by
\begin{equation}\label{eq:Z}
    \zso \leftarrow \zso \sqrt{ \frac{\sum_{f,t} \hat{y}_{fts} \text{ }tr(\Xe^{-1}\X\Xe^{-1}\Wfo)}{\sum_{f,t} \hat{y}_{fts} \text{ } tr(\Xe^{-1}\Wfo)}}\text{.}
\end{equation}
\noindent where $tr(\mathbf{X})= \sum^M_{m=1} x_{mm}$ is the trace of a square matrix $\mathbf{X}$.
Finally, update rules for the DoA Kernels $\Wfo$ are obtained by solving an algebraic Ricatti equation. For the sake of brevity, its computation has been omitted here, regardless refer to~\cite{Sawada2013} for more details.

\subsection{Source separation using the learned spatial properties}\label{sec:refinement}
Once the mixing filter is estimated, we propose to refine the singing voice spectrogram using the previously estimated mixing filter. The aim is to mitigate the interference caused by the other sources using the spatial properties learned from the multichannel mixture.
In other to refine the time frequency spectrogram $\hat{y}_{fts}$ obtained from the MaDTwinNet prediction, we further decompose it into a linear combination of spectral patterns and time-varying gains. To this end, we used the high performance Hierarchical Alternating Least Squares Algorithm (HALS) approach~\cite{Cichocki2009a}. The HALS updating rules which minimize the IS divergence are defined as,
\begin{align}
    % &b_{fks} \leftarrow \frac{1}{\sum_t \gkts} \sum_t \left[\hat{y}_{fts}\right]_+\text{and}\\
    &b_{fks} \leftarrow (\sum_t \gkts)^{-1}\sum_t \left[\hat{y}_{fts}\right]_+\text{and}\\
% \end{equation}
% \begin{equation}
%   &\gkts \leftarrow \frac{1}{\sum_f  b_{fks}} \sum_f  \left[ \hat{y}_{fts}\right]_+\text{,}
  &\gkts \leftarrow(\sum_f  b_{fks})^{-1}\sum_f  \left[ \hat{y}_{fts}\right]_+\text{,}
\end{align}
\noindent where $[\cdot]_+$ denotes $max\{\epsilon,\cdot\}$ with $\epsilon$ a small constant to enforce positive entries, $b_{fks}\in\mathbb{R}_{\geq0}$ refers to the spectral basis patterns and $g_{kts}\in\mathbb{R}_{\geq0}$ is their corresponding time-varying gains.
Then, both parameters $b_{fks}$ and $\gkts$ are refined with CNMF using the spatial information of the estimated mixing filter. In this way, the SCM signal model can be expressed as,
\begin{equation}\label{eq:signalmodel}
    \X \approx \Xe = \sum_{s=1}^S \sum_{k=1}^K \Hfs \underbrace{b_{fks} \gkts}_{\hat{y}_{fts}}\text{.}
\end{equation}

Note that the mixing filter in Eq.~\eqref{eq:signalmodel} is fixed from the previous step while the basis functions and time varying gains are the free parameters initialized from the MaDTwinNet estimation using the HALS method.
The multiplicative update rules for the free parameters in the signal model in Eq.~\eqref{eq:signalmodel} for Itakura Saito divergence are defined as follows~\cite{Sawada2013},
\begin{align}
    &b_{fks} \leftarrow b_{fks}  \sqrt{\frac{\sum_{f}\gkts \text{ }tr(\Xe^{-1}\X\Xe^{-1}\Hfs)}{\sum_{f}\gkts \text{ }tr(\Xe^{-1}\Hfs)}}\text{ and}\label{eq:B}\\
% \end{equation}
% \begin{equation}
    &\gkts \leftarrow \gkts \sqrt{\frac{\sum_{f} b_{fks} \text{ }tr(\Xe^{-1}\X\Xe^{-1}\Hfs)}{\sum_{f} b_{fks} \text{ }tr(\Xe^{-1}\Hfs)}}\text{.}\label{eq:G}
\end{align}

Once the model parameters have been optimized, the reconstruction of the source signals is performed using a generalized Wiener filtering strategy,% where the estimated CNMF magnitude spectrogram for each sound source $s$ and the generalized Wiener mask are computed as
% where the estimated CNMF magnitude spectrogram for each sound source $s$ can be expressed as
%\begin{align}
%    \bar{y}_{msft} &= \sum_o tr(\mathbf{W}_{fo})_m \zso \sum_k b_{fks} \gkts\text{ and}\\
%    \tilde{{y}}_{msft} &= \frac{\tilde{{x}}_{msft}\cdot \bar{y}_{msft}}{\sum_{s,o} tr(\mathbf{W}_{fo})_m z_{so} \sum_k b_{fks} \gkts}\text{,}
%\end{align}
\begin{equation}
    \tilde{y}_{mfts} = \frac{\sum_o tr(\mathbf{W}_{fo})_m \zso \sum_k b_{fks} \gkts}{\sum_{s,o} tr(\mathbf{W}_{fo})_m z_{so} \sum_k b_{fks} \gkts} \cdot \tilde{x}_{mft}.
\end{equation}
\noindent where $\tilde{x}_{mft} \in \mathbb{C}$ is the input multichannel signal mixture time-frequency spectrogram. Finally, the multichannel time-domain signals are obtained by the inverse STFT of $\tilde{y}_{mfts} \in \mathbb{C}$ and frames are combined by weighted overlap-add.

%%%%%%%%%%%%%%%%%%%%%%%%%%%%%%%%%%%%%%%%%%%%%%%%%%%%%%%%%%%%%
%%%%%%%%%%%%%%%%%%%%%%%%%%%%%%%%%%%%%%%%%%%%%%%%%%%%%%%%%%%%%
\section{Evaluation and results}\label{sec:evaluation}
\subsection{Experimental Setup}\label{sec:experimental}
In this work, the performance of our proposed framework is assessed by focusing on the task of singing voice separation. The experimental evaluation was carried out on the Demixing Secret Dataset\footnote{\url{https://sigsep.github.io/datasets/dsd100.html}} (DSD100). A subset of 50 mixtures has been used for the training process together with their corresponding individual sources, and another 50 mixtures for the evaluation.
% \begin{figure*}[!t]
%     \centering
%     \begin{subfigure}[b]{0.35\textwidth}
%     \centering
%         \includegraphics[width=\textwidth]{BSS_5cm.eps}
%     \end{subfigure}
%     ~
%     \begin{subfigure}[b]{0.35\textwidth}
%     \centering
%         \includegraphics[width=\textwidth]{BSS_1m.eps}
%     \end{subfigure}
%     \caption{Singing voice separation results using Itakura Saito divergence for the proposed dataset with two different microphone arrays (small and large array). Boxes denote 25\% and 75\% quartiles and whiskers the extreme points.}
%     \label{fig:results}
% \end{figure*}
\begin{comment}
\begin{figure}[!t]
    \centering
    \begin{subfigure}[b]{.9\columnwidth}
    \centering
        \includegraphics[width=.75\textwidth]{BSS_5cm.eps}
    \end{subfigure}
    ~
    \begin{subfigure}[b]{.9\columnwidth}
    \centering
        \includegraphics[width=.75\textwidth]{BSS_1m.eps}
    \end{subfigure}
    \caption{Singing voice separation results using Itakura Saito divergence for the proposed dataset with two different microphone arrays (small and large array). Boxes denote 25\% and 75\% quartiles and whiskers the extreme points.}
    \label{fig:results}
\end{figure}
\end{comment}
The source material consists of a set of four source signals: ``accompaniment'', ``bass'', ``drums'' and ``vocals''. This material was recorded in isolation. Therefore, we simulated different mixing conditions using the Roomsim Toolbox \cite{Campbell2005} for a rectangular room of dimensions $7 m \times 12 m \times 3 m$ and omnidirectional microphones. The average reverberation time is $T_{60} = 650 ms$, which provides a moderate reverberation environment.
%which provides a typical concert hall acoustic environment. 
We considered two two-channel microphone arrays varying the inter-microphone distance: 5 cm (small array) and 1 m (large array). The four sources have been mixed at directions $(\phi,\theta) = [(0^\circ,0^\circ), (45^\circ,0^\circ),(90^\circ,0^\circ),(135^\circ,0^\circ)$ in azimuth and elevation and 1 m of distance with respect to the center of the microphone pair.

The evaluation of our proposal was carried out with the BSS\_EVAL toolbox \cite{Emiya2011a} which implements three metrics: \emph{Source to Distortion Ratio} (SDR), \emph{Source to Interference Ratio} (SIR) and \emph{Source to Artifacts Ratio} (SAR). These metrics are widely accepted in the field of source separation and thus, allow a fair comparison with other state-of-the-art methods. For our method, the time-frequency representation is obtained from Short-Time Fourier Transform (STFT) using 2048-point FFT and half-window overlap. The maximum number of iterations for the CNMF decomposition is set to 200, and number of look directions $O$ to 60. 

Different configurations of the proposed separation framework has been compared in order to show its separation capabilities. First, \textit{CNMF Fix} denotes the proposed method using the spectral patterns learned from the output of the MaD TwinNet and keeping them fixed during the CNMF stage (i.e., the parameters $\mathbf{W}_{fo}$, $z_{so}$ and $\gkts$ are updated  by solving the Ricatti equation, \eqref{eq:Z} and \eqref{eq:G}, respectively). \textit{CNMF Free} denotes the proposed method using the spectral patterns learned from the output of the MaD TwinNet and updating them during the factorization (i.e., the parameters $\mathbf{W}_{fo}$, $z_{so}$, $b_{fks}$ and $\gkts$ are updated  by solving the Ricatti equation, \eqref{eq:Z}, \eqref{eq:B} and \eqref{eq:G}, respectively). 
%First, \textit{CNMF Fix} denotes the proposed method using the spectral patterns learned from the output of the MaD TwinNet and keeping them fixed during the CNMF stage. \textit{CNMF Free} denotes the proposed method using the spectral patterns learned from the output of the MaD TwinNet and updating them during the factorization. 
%
For both cases, we have chosen the separated MaD TwinNet signal from the channel where the singing voice signal is predominant. \textit{Oracle} denotes the proposed method using the spectral patterns learned from the reverberant source images at the microphones, which provides the ground-truth for the singing voice source.
%\textit{Oracle} denotes the proposed method using the spectral patterns learned from the dry signals from the database, which provides the ground-truth for the singing voice source. 
This variant represents the upper bound for the best separation that can be reached with the proposed framework. Finally, \textit{Rand} denotes the proposed method initializing the basis functions to random values. This evaluation provides a starting point for the separation algorithms.
% \begin{table}[htbp]
%   \centering
%     \begin{tabular}{lrrrrrr}       & \multicolumn{3}{c}{Small array} & \multicolumn{3}{c}{Large array} \\ \noalign{\smallskip}
% \cline{2-7}     \noalign{\smallskip}      & \multicolumn{1}{c}{SDR} & \multicolumn{1}{c}{SIR} & \multicolumn{1}{c}{SAR} & \multicolumn{1}{c}{SDR} & \multicolumn{1}{c}{SIR} & \multicolumn{1}{c}{SAR} \\ \noalign{\smallskip}
%     \hline \noalign{\smallskip}
%     Random   & -0.77 & -6.06 & 5.45  & -0.85 & -9.11 & 4.79 \\ \noalign{\smallskip}
%     Oracle  & 5.44  & 15.52 & 7.25  & 5.15  & 13.78 & 7.44 \\  \noalign{\smallskip} \hline \noalign{\smallskip}
%     MaD TwinNet  & 2.89  & 9.26  & 2.71  & 2.96  & 9.35  & 2.83 \\ \noalign{\smallskip}
%     MNMF     & 3.65  & 6.96  & 6.02  & 3.93  & 7.09  & 6.49 \\ \noalign{\smallskip}
%     CNMF Fix     & 3.57  & \textbf{14.09} & 5.60  & 4.35  & \textbf{12.36} & 6.56 \\ \noalign{\smallskip}
%     CNMF Free    & \textbf{5.22} & 12.99 & \textbf{6.97} & \textbf{5.01} & 11.40 & \textbf{7.27} \\
%      \noalign{\smallskip}\hline
%     \end{tabular}
%     \caption{Median values for SDR, SIR and SAR of the proposed variants and baseline methods.}
%   \label{tab:medianValues}
% \end{table}

We also compare our proposal with a supervised multichannel NMF-based method~\cite{Miron2016} (\textit{MNMF}). This method models the multichannel audio magnitude spectrogram using NMF with a mixing panning matrix which encodes the contribution of each source to each channel. As in our \textit{CNMF Free} approach and for a fair comparison, this method uses as basis functions the same spectral patterns learned in advance from the output of the MaD TwinNet, and updates them during the factorization. Its time-varying gain matrix is initialized to random values. Moreover, in order to show the improvement of including the CNMF stage, the output of the MaD TwinNet system has been evaluated performing the single-channel separation over both channels.
\vspace{-3pt}
\begin{figure}[!t]
\centering
\includegraphics[width=0.9\columnwidth]{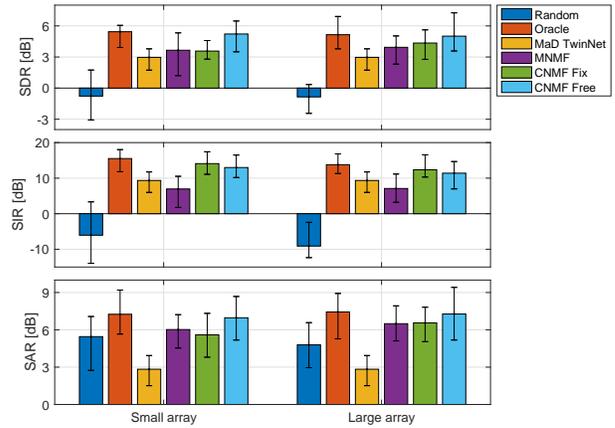}
\caption{Singing voice separation results using Itakura Saito divergence for the proposed dataset with two different microphone arrays (small and large array). 
% Boxes denote 25\% and 75\% quartiles and whiskers the extreme points.
}
\label{fig:results}
\end{figure}
\subsection{Results}
% \begin{figure}
%     \centering
%     \begin{subfigure}[b]{0.7\columnwidth}
%     \centering
%         \includegraphics[width=\columnwidth]{BSS_5cm.eps}
%     \end{subfigure}
%     ~
%     \begin{subfigure}[b]{0.7\columnwidth}
%     \centering
%         \includegraphics[width=\columnwidth]{BSS_1m.eps}
%     \end{subfigure}
%     \caption{Singing voice separation results using Itakura Saito divergence for the proposed dataset with two different microphone arrays (small array in the left column and large array in rigth column). Boxes denote 25\% and 75\% quartiles and whiskers the extreme points.}
%     \label{fig:results}
% \end{figure}
%
The results obtained by the configuration and methods introduced above are illustrated in Figure~\ref{fig:results}, 
% For a better visualization of the results, the median values obtained in the evaluation are summarized in Table~\ref{tab:medianValues}.
% As can be observed, 
where there is a similarity on the values between the short and large arrays. 
% where the results show a similar behaviour for the short and large arrays. 
Random variant obtains the worst results, which seems logical since no prior information about the target signal is provided. On the other hand, the best results are reached for the Oracle version and inform us about the best separation that can be achieved using the proposed approach with an optimal set of singing voice spectral patterns. Regarding the baseline methods, our proposed framework clearly outperforms them in terms of SDR, SIR and SAR. Note that the improvement of the CNMF Free variant over MNMF is due to the fact that the phase information can be modeled with CNMF algorithm. The underperformance of MaD TwinNet is because it is a single-channel algorithm and does not take into account inter-channel source information.

Finally, CNMF Fix is around 1.7 dB below CNMF Free in SDR, which means that tuning the basis functions during the factorization process further improves separation. This is due to the localization estimation which helps to refine the spectral patterns of the singing voice mitigating the problem of the overlapping partials, and therefore, improving the prediction of the source spectrograms. In fact, we can see that CNMF Free achieves a result very close to the Oracle variant.

\textcolor{black}{In relation to subjective results, some listening demos can be found at the web page of results\footnote{\url{https://antoniojmm.github.io/MultichannelSingingVoice_MadTwin-CNMF/}}.}

\section{Conclusions}
In this paper, we present a multichannel singing voice separation framework using a SCM-based CNMF signal model informed using Deep Learning. In particular, the proposed framework consists of two steps: mixing filter estimation and source estimation refinement. In the first step, a DoA Kernel based mixing filter is estimated using information from the  magnitude spectrograms of the  sources, obtained using the single channel MaDTwinNet method. Then, in the second step, the estimated source time-frequency magnitude spectrograms are refined using the spatial information learned in the first step. This information allows to reduce the interference caused by correlated sources arriving from different spatial directions. The results show improved separation performance in comparison to state-of-the-art methods. Specifically, we reached an increase of 1.5 dB and 6 dB in SDR and SIR, respectively

As future work, we would investigate the integration of the mixing filter output within the network architecture. Additionally, we will investigate the computational optimization of the CNMF stages.
\bibliographystyle{unsrt}
\bibliography{references}
\end{document}